\documentclass{article}
\usepackage{frascatiphys_SP}
\begin{document}
\title{
FINAL--STATE RADIATION IN ELECTRON--POSITRON ANNIHILATION INTO
PION PAIR}
\author{
O~. Shekhovtsova \thanks {\hspace{0.2cm}  co--autors: S.~Dubinsky,
A.~Korchin and N.~Merenkov -- \em NCS KIPT, Ukraine;
\em G.~Pancheri-- \em INFN Labaroratori Nazionale di Frascati, Italy}\\
\em NCS KIPT, Akademicheskaya 1, Kharkov 61108, Ukraine \\
\em shekhovtsova@kipt.kharkov.ua} \maketitle \baselineskip=11.6pt
\baselineskip=14pt

\vspace{0.5cm}
\section{Introduction}
In this article the final-state radiation (FSR)  of the hard
photon in $e^-(p_1)+e^+(p_2)\to\gamma^*(Q)\to\pi^+(p_+)+
\pi^-(p_-)+\gamma(k)$ reaction is considered in the framework of
ChPT with vector $\rho$ and axial--vector $a_1$ mesons
\cite{Ecker_89} (the FSR diagrams are shown in Fig.1).

Our consideration of FSR is motivated by the necessity to study
model dependence of the next-to-leading order hadronic
contribution $a_\mu^{had,\gamma}$ to anomalous magnetic moment
(AMM) of the muon ($ a_\mu^{had, \gamma}$ is the hadronic
contribution, where additional photon is attached to hadrons).
Also FSR is a main unrestricted background  to scan the hadronic
cross--section at meson factories by the radiative return method
\cite{KLOE}. In this method only ISR (initial-state radiation)
events have to be chosen and the FSR processes have to be
rejected. Different methods have been suggested to separate ISR
and FSR contributions for the dominant hadronic channel at low
energies -- the pion-pair production. One of them is to choose
kinematics, where photon is radiated outside the narrow cones
along the momenta of the pions. In these conditions the FSR
contribution is suppressed.  If the FSR background can be reliably
calculated in some theoretical model then it can be subtracted
from experimental cross section of $e^+ e^- \to \pi^+ \pi^-
\gamma$ or incorporated in the Monte Carlo event generator used in
analysis. Finally, the theoretical predictions for FSR can be
tested by studying the $C$--odd interference of ISR and FSR
\cite{Czyz_03}.

The FSR cross section  has been calculated \cite{
Czyz_03} in framework of the scalar QED (sQED), in which the pions
are treated as point-like particles, and the resulting amplitude
is multiplied by the pion electromagnetic form factor $F_\pi (s)$
evaluated in VMD model ($s$ is the total $e^+e^-$ energy
squared) to account for the pion structure. 
Although sQED in some cases works well \cite{KLOE,Czyz_03}, it is
clear that sQED is a simplified model of a complicated process,
which may include excitation of resonances, loop contributions,
etc. In view of the high requirements for the accuracy of
theoretical predictions for AMM, further studies of the FSR
contribution are necessary.
\section{Results of calculation}
\label{sec:results}
\begin{figure}
\begin{center}
\vspace{5cm} \includegraphics{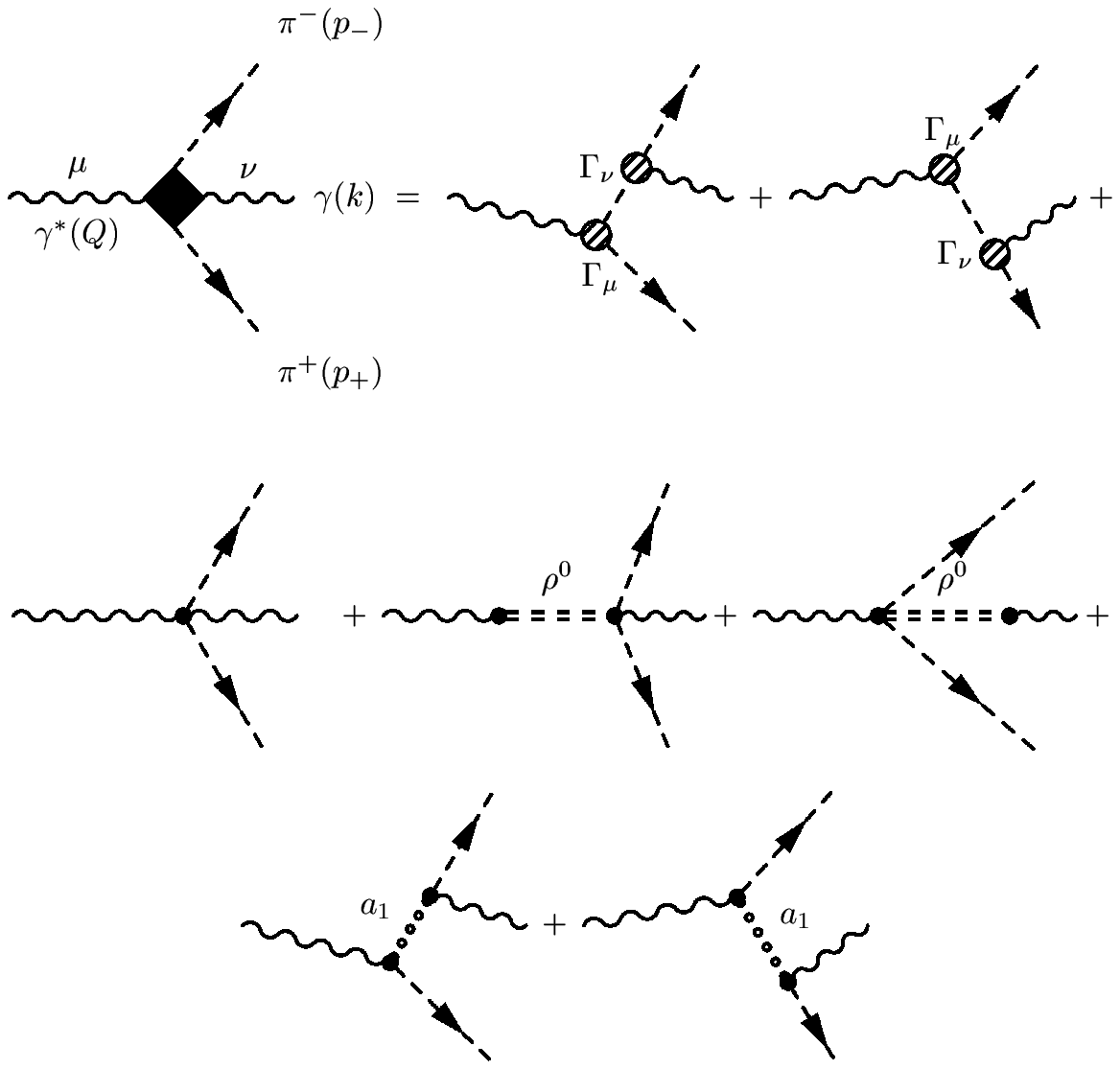}
 \label{fig2}
\end{center}
\caption{Diagrams for FSR in the framework of ChPT.}
\end{figure}
In view of the restricted space of this contribution only the
results of calculations are presented (for details see Ref.
\cite{Dub_04}).

First, the charge asymmetry 
\cite{Czyz_03} proportional to the interference of ISR and FSR
 is calculated for the so-called collinear kinematics in which
 the hard photon is radiated inside a narrow cone with the
 opening angle $2\theta _0$ \ ($\theta_0\ll 1$)
along the direction of initial electron. In Fig.1 we show the
asymmetry dependence on pion polar angle at fixed two--pion
invariant mass $q^2$. It follows that the asymmetry changes sign
at about $q^2=0.5$ GeV$^2$. At all pion angles the difference
between sQED and ChPT shows up only at small values of $q^2$ or,
equivalently, at high photon energies. Thus only at high photon
energies the contribution from $a_1$ intermediate meson (see
diagrams with $a_1$--meson in Fig.2) is sizable. For large values
of $q^2$ the difference between predictions of sQED and full
calculation in ChPT is small: for $q^2\geq 0.6$ GeV$^2$ it is less
than $1\%$ (the dashed and solid lines almost coincide in Fig.1).
Taking into account that the asymmetry itself is less than
$10^{-2}$, the experimental observation of such deviations in the
energy region $q^2\geq 0.6$ GeV$^2$ is problematic.

\begin{figure}[t]
 \vspace{4cm}
\includegraphics{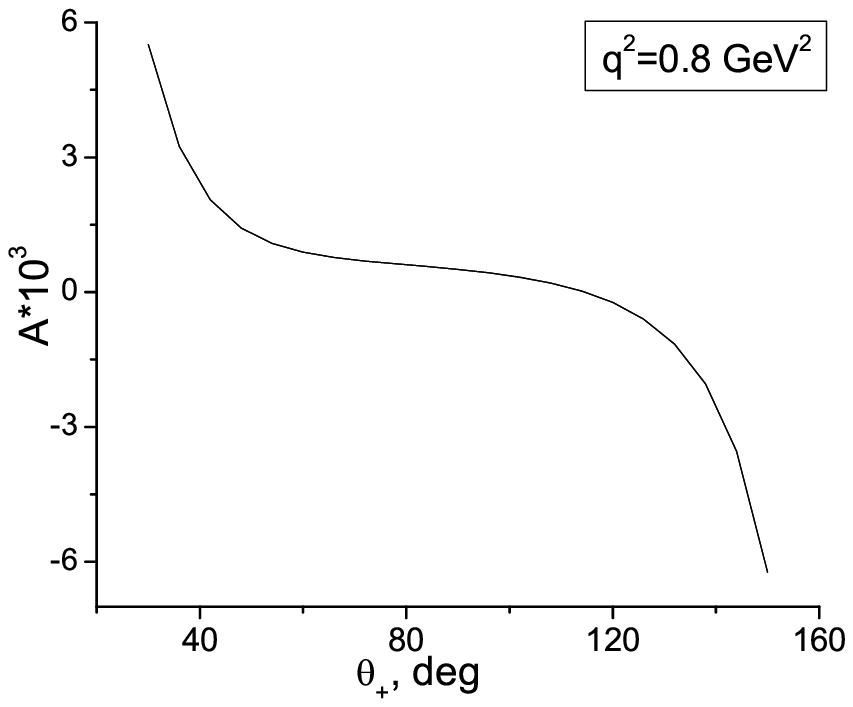}
\includegraphics{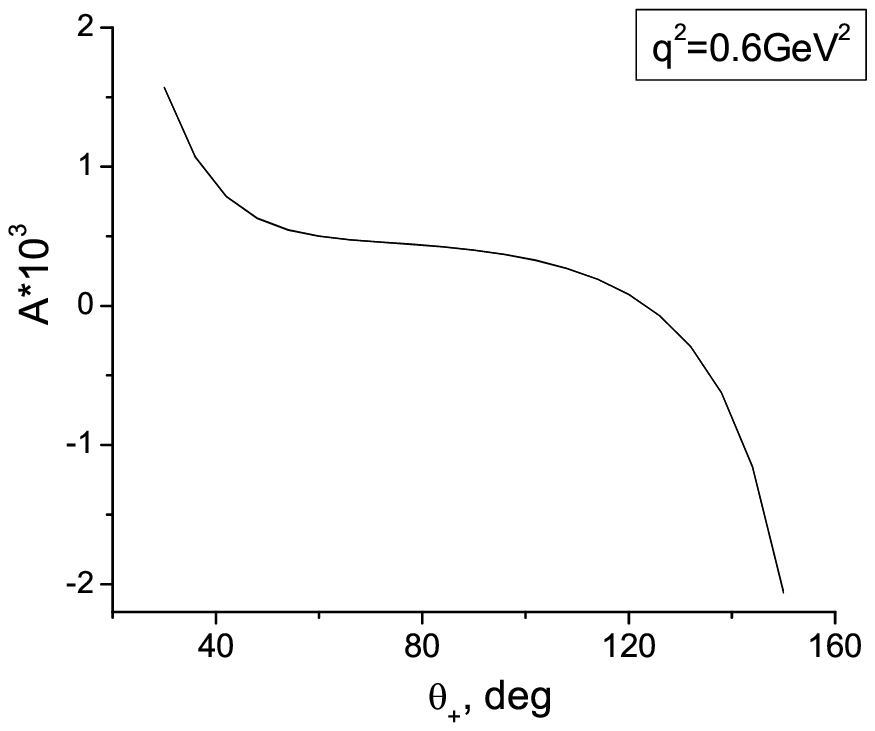}
\includegraphics{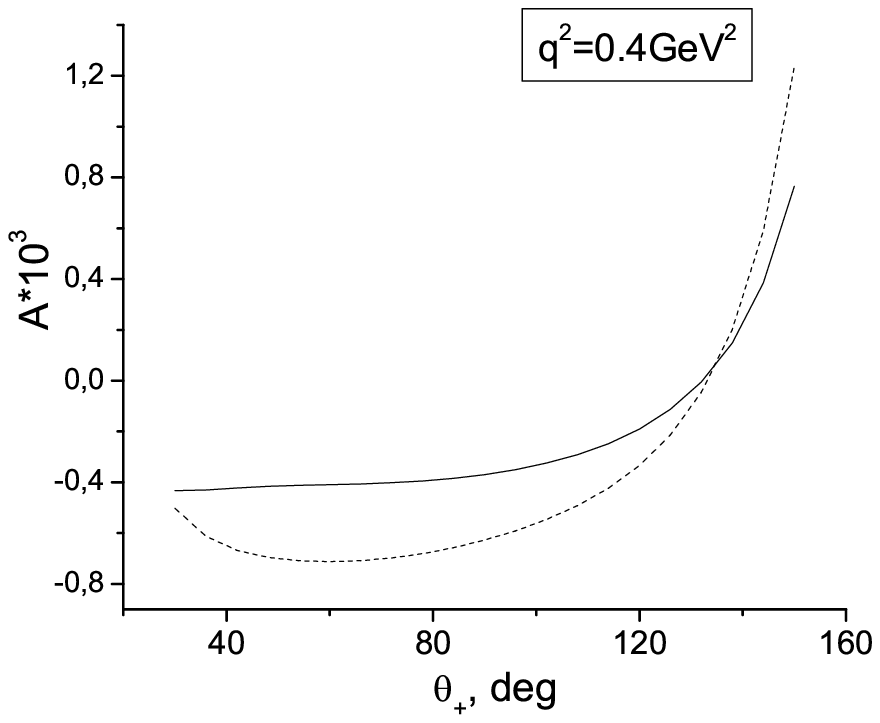}
         \vspace{-1.5cm}
 \caption{\it
      Charge asymmetry as a function of pion
polar angle at fixed $q^2$ for $s=1$ GeV$^2$. The solid line
corresponds to sQED, the dashed line -- the full result in ChPT.
    \label{fig_as} }
\end{figure}
\begin{figure}[t]
 \vspace{4cm}
\includegraphics{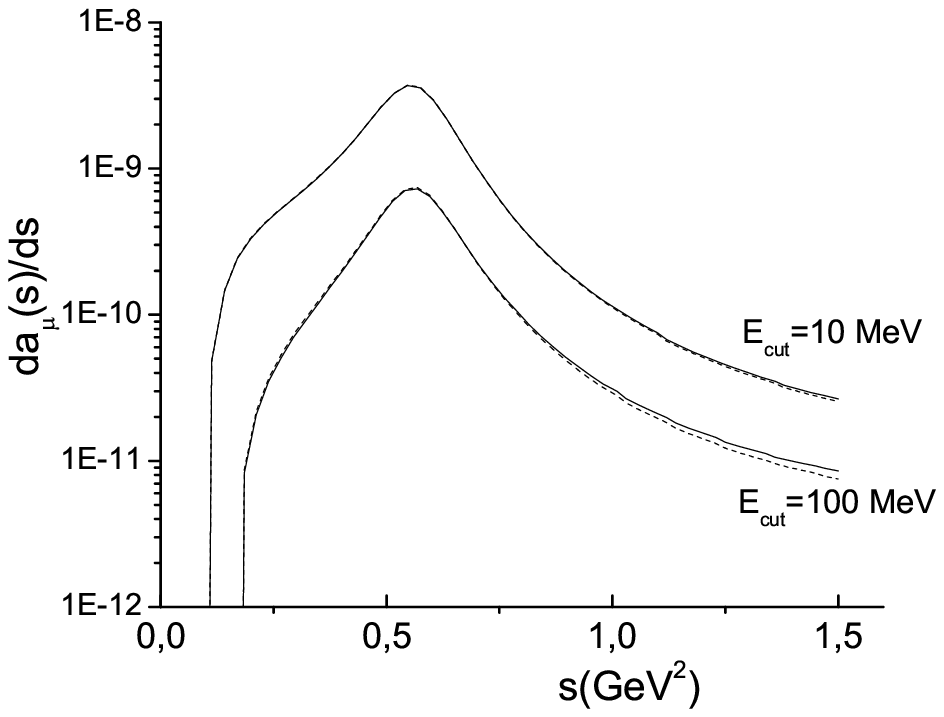}
\includegraphics{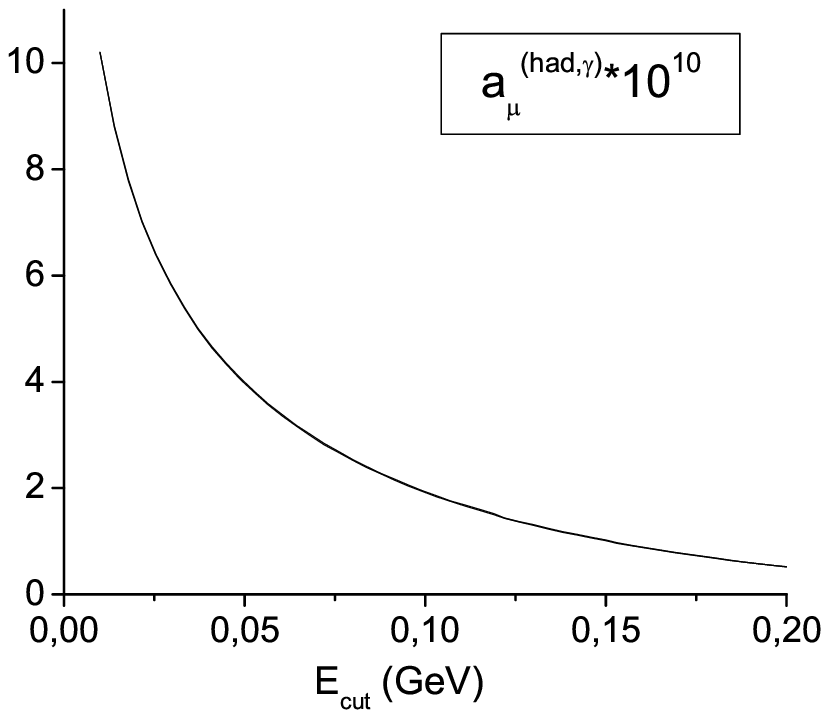}
 \includegraphics{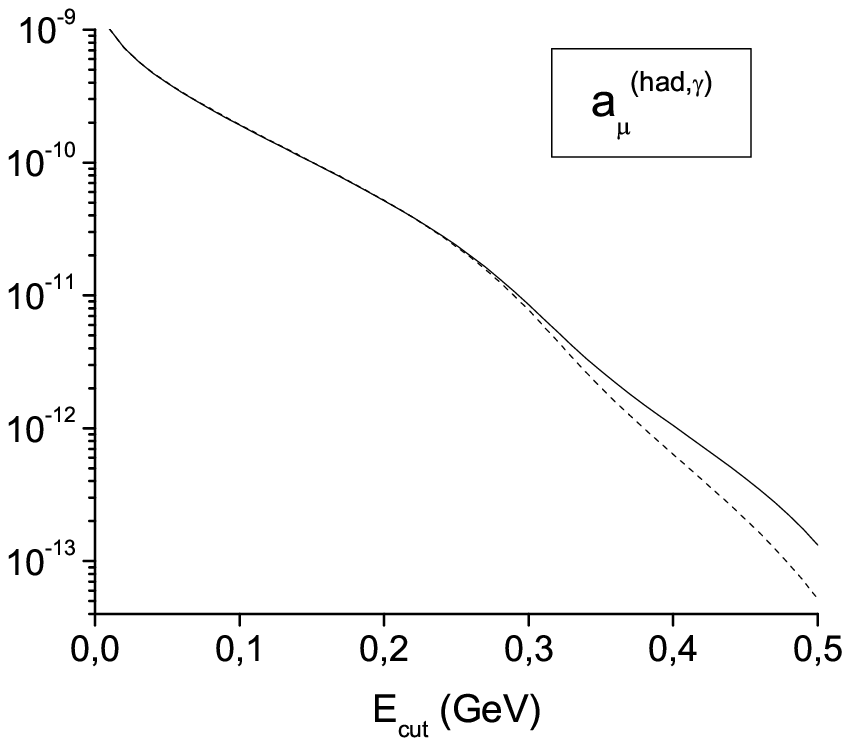}
        \vspace{-.2cm}
 \caption{\it
      Differential contribution
$a_\mu^{\pi\pi,\gamma}$
 (left panel). Integrated contribution to
$a_\mu^{\pi\pi,\gamma}$ as a function of $E_{cut}$ (central and
right panels). Here $s_{max}=1.5$ GeV$^2$. Notations for the
curves are the same as in Fig.2.
    \label{muon_anom} }
\end{figure}
Second, we apply the result of Ref. \cite{Dub_04} to evaluation of
$a_\mu^{\pi\pi,\gamma}$. It appears that the additional
contributions to $a_\mu^{\pi\pi,\gamma}$ arising in ChPT are very
small compared with sQED result (here only the radiation from hard
photon ($\omega\geq E_{cut}$) is taken into account). Even for
$E_{cut}=200$ MeV the ChPT result differs from the sQED one by
only $3.5\%$ (see the solid and dashed lines in Fig.3 which almost
coincide). These small deviations are not surprising. First, at
fixed value of $s$ the low--energy photon region, which is
described in a similar way by both models, dominates in
$a_\mu^{\pi\pi,\gamma}$. Second, the main contribution to
$a_\mu^{\pi\pi,\gamma}$  comes from the region of the
$\rho$--resonance, which is treated in the same manner in sQED and
ChPT via VMD model.

At the same time, with increasing the photon energy sQED losses
its predictive power. This is demonstrated in Figs.2 and 3 (right
panel). In this region the contribution from $a_1$--meson is
considerable and has to be taken into account. For example, at the
photon energy about 500 MeV the deviation from sQED reaches
$60\%$. However, this deviations (which are of the order of
$10^{-12}$) are beyond the accuracy of the present measurements of
the muon AMM.

\section{Conclusions}\label{sec:conclusions}

We demonstrated  that the model dependence of the two--pion
contribution to $a_\mu^{had,\gamma}$ is weak, and the value of
$a_\mu^{had,\gamma}$ is not sensitive to chiral dynamics beyond
the $\rho$--meson dominance. As for the charge asymmetry, its
model dependence can be observed experimentally only for $q^2$
near the two-pion threshold region: $4m_\pi^2\leq q^2<0.4$
GeV$^2$.

Therefore, in the bulk of energies up to $1$ GeV, sQED is
sufficient to describe the FSR contribution to both
$a_\mu^{had,\gamma}$ and $C$--odd asymmetry. To observe deviations
from sQED the existing experimental error bars for
$a_\mu^{had,\gamma}$ have to be reduced by at least one order of
magnitude. Possibly, the more complicated many--particle channels
in $e^+ e^-$ annihilation are more sensitive to the chiral
dynamics.

\end{document}